\documentclass[showpacs,amssymb,amsmath,nobibnotes, aps,prl,secnumarabic,superscriptaddress,twocolumn]{revtex4}

\usepackage{bm}
\usepackage{natbib}
\usepackage{graphicx}

\begin{document}
\textfloatsep 10pt

\title{What drives mesoscale atmospheric turbulence?}
\author{H. Xia}
\author{H. Punzmann}
\affiliation{Research School of
Physical Sciences and Engineering,\\ Australian National
University, Canberra ACT 0200, Australia}
\author{G. Falkovich}
\affiliation{Physics of Complex Systems, Weizmann Institute of
Science,  Rehovot 76100, Israel}

\author{M.G. Shats}
\email{Michael.Shats@anu.edu.au}
\affiliation{Research School of
Physical Sciences and Engineering,\\ Australian National
University, Canberra ACT 0200, Australia}

\date{\today}

\begin{abstract}
Measurements of atmospheric winds in the mesoscale range (10-500
km) reveal remarkably universal spectra with the $k^{-5/3}$ power
law. Despite initial expectations of the inverse energy cascade,
as in two-dimensional (2D) turbulence, measurements of the third
velocity moment in atmosphere, suggested a direct energy cascade.
Here we propose a possible solution to this controversy by
accounting for the presence of a large-scale coherent flow, or a
spectral condensate. We present new experimental laboratory data
and show that the presence of a large-scale shear flow modifies
the third-order velocity moment in spectrally condensed 2D
turbulence, making it, in some conditions, similar to that
observed in the atmosphere.
\end{abstract}

\pacs{47.27.-i, 47.27.Rc, 47.55.Hd, 42.68.Bz}

\maketitle

Atmospheric motions are powered by gradients of solar heating.
Vertical gradients cause thermal convection on the scale of the
troposphere depth (less than 10 km). Horizontal gradients excite
motions on a planetary (10000 km) and smaller scales. Both inputs
are redistributed over wide spectral intervals by nonlinear
interactions \cite{Kraichnan67,Lilly83,Gage_Nastrom86}. The
spectra of kinetic energy of atmospheric winds have been analyzed
during the Global Atmospheric Sampling Program \cite{NG84}. These
wavenumber spectra measured in the upper troposphere and in the
lower stratosphere have shown two power laws: $E(k) \propto
k^{-5/3}$ for the scales between 10 and 500 km, and a steeper
spectrum with $E(k) \propto k^{-3}$ in the range of scales
(500-3000) km (similar to the spectra in Figs. 2,3). According to
the Kraichnan theory of 2D turbulence \cite{Kraichnan67},
$E(k)=C\epsilon ^{2/3}k^{-5/3}$ corresponds to an inverse energy
cascade and $E(k)=C_q \eta ^{2/3}k^{-3}$ to a direct vorticity
cascade. Here $\epsilon$ and $\eta$  are the dissipation rates of
energy and enstrophy respectively. This prompted a two-source
picture of atmospheric turbulence with a planetary-scale source of
vorticity and depth-scale source of energy \cite{Lilly89}. The
large-scale part of the spectrum can be due to a direct vorticity
cascade \cite{Lilly89} or it can result from an inverse cascade of
inertio-gravity waves \cite{Falkovich92}. It can also result from
the energy pile-up at the system scale in the process of spectral
condensation \cite{Smith_Yakhot_94} (a turbulent counterpart of
Bose condensation), or it can be due to a combination of the
above. What about the mesoscale 5/3-spectrum? Is it an energy
cascade and what is the flux direction?

In homogeneous turbulence, spectral energy flux is expressed via
the third-order moment of the velocity \cite{Monin_Yaglom,
Lindborg99} $S_3= \langle (\delta V_L)^3 \rangle + \langle \delta
V_L (\delta V_T)^2 \rangle$ . Here $\delta V_L$ and $\delta V_T$
denote the difference of velocities at two points separated by
distance $r$. Angular brackets denote ensemble averaging over
realizations, and the subscripts denote the longitudinal ($L$) and
transverse ($T$) velocity components relative to $r$. Positive
$S_3$ corresponds to the inverse energy cascade from small to
large scales. Measurements of the third moment of the velocity
difference in the atmosphere gave a negative value in the interval
10-100 km, which was interpreted as the signature of the forward
cascade \cite{Cho_Lindborg01}. The negativity of the third moment
spawned hypotheses about a direct energy cascade in 2D or
stratified turbulence \cite{Gkioulekas06, Brethouwer07}. Here we
demonstrate experimentally that a negative small-scale $S_3$ in a
system with an inverse cascade can be caused by a large-scale
shear flow.

Let us first consider how small- and large-scale parts of the velocity difference (respectively $\delta v$ and $\delta V$) contribute to the second and third velocity moments. Comparing $\langle(\delta V)^2 \rangle \cong s^2r^2$ with $\langle(\delta v)^2 \rangle \cong C(\epsilon r)^{2/3}$ we see that the small-scale (turbulent) part dominates at the scales smaller than $l_t \cong C^{3/4}s^{-3/2} \epsilon ^{1/2}$. Here $s=V/L_s$ is a large-scale velocity gradient and $L_s$ is the velocity shear scale length which depends on the system size and on the topology of the large-scale flow. For the third moment, we compare $\langle(\delta v)^3 \rangle \cong \epsilon r$ with the cross-correlation term $\langle \delta V(\delta v)^2 \rangle \cong srC(\epsilon r)^{2/3}$ and observe that the influence of $\delta V$ extends to a much smaller scale $l_* \cong C^{-3/2} s^{-3/2} \epsilon ^{1/2}$, because the dimensionless constant $C$ is typically larger than unity, as discussed below.

The above estimates are true for a large-scale part produced by any source. In particular, when it is produced as a condensate by an inverse cascade \cite{Smith_Yakhot_93,Molenaar04,Chertkov07,Sommeria86,Paret_Tabeling_98,Shats07} (as in the experiments described below) one estimates $s$ as follows. Let the linear damping rate $\alpha$  be smaller than the inverse turn-over time $C^{1/2} \epsilon ^{1/3} L^{-3/2} $ for the vortices comparable to the system size $L$. Then the flow coherent over the system size (the condensate) appears \cite{Smith_Yakhot_93,Molenaar04,Chertkov07,Sommeria86,Paret_Tabeling_98,Shats07} with the velocity estimated from the energy balance, $\alpha V^2 \approx 2\epsilon$, which gives  $s \cong V/L_s \cong L^{-1}_s \sqrt{2\epsilon / \alpha}$ and

\begin{equation}\label{Eq:k_t}
 k_t = \pi / l_t \cong \pi L_s^{-3/2} (C \alpha /2)^{-3/4} \epsilon ^{1/4}.
\end{equation}
\noindent Note that this is not the condition that the turnover time at $l_t$ is $\alpha ^{-1}$, as in \cite{Smith_Yakhot_93}; incidentally Eq.~\ref{Eq:k_t} gives a correct estimate ($k_t \cong 1$) for their conditions. The spectrum $E(k) \propto k^{-3}$ at $k<k_t$ is due to the condensate \cite{Smith_Yakhot_93, Shats07, Chertkov07}, while $E(k)=C\epsilon ^{2/3}k^{-5/3}$ is expected at $k_t<k<k_f$.

\begin{figure}
\includegraphics[width=8.5 cm]{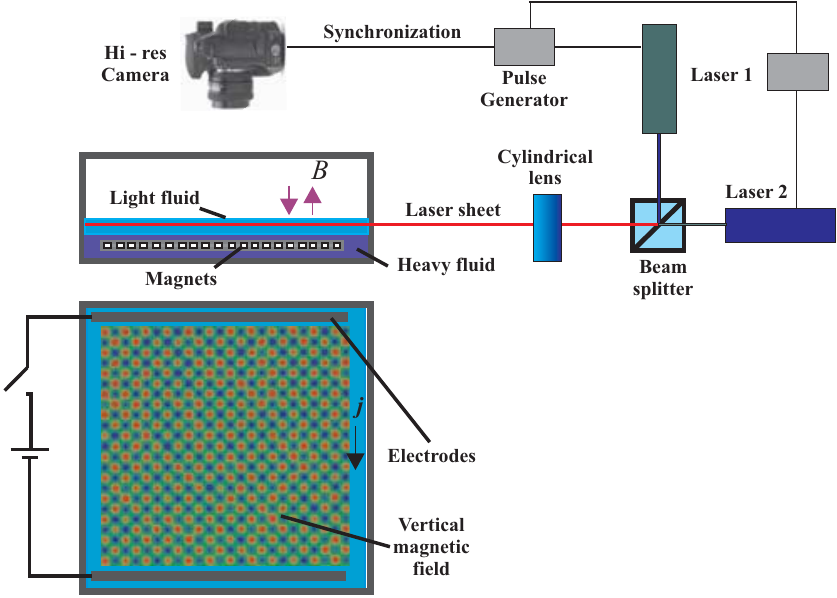}
\caption{\label{fig_0} Schematic of the experimental setup}
\end{figure}

Here we report the experiment in which the strength and the spectral extent of the condensate are varied by changing either $\alpha$ or $L$. The experimental setup shown in Figure~\ref{fig_0} is similar to those described in \cite{Paret_Tabeling_98, Shats07, Chen06} but has a substantially larger number of forcing vortices (up to 900), higher spatial resolution and larger scale separation ($L/l_f \approx 30$). A turbulent flow is generated electromagnetically in stratified thin fluid layers whose thicknesses are varied to achieve different damping rates $\alpha$. A heavier non-conducting fluid (Fluorinert, specific gravity SG =1.8) is placed at the bottom. A lighter conducting fluid, NaCl water solution (SG =1.03), is placed on top. Non-uniform magnetic field is produced by a square matrix of $30 \times 30$ permanent magnets (10 mm apart). The electric current flowing through the top (conducting) layer produces 900 $J \times B$-driven vortices which generate turbulence. Square boundaries with $L = (0.09-0.24)$ m are used. To visualize the flow, imaging particles (polyamid, $50 \mu m$, SG= 1.03) are suspended in the top layer and are illuminated by a 1 mm laser sheet parallel to the fluid surface. Laser light scattered by the particles is filmed from above using a 12 Mpixel camera. Green and blue lasers ($\lambda = 532$ nm and $\lambda = 473$ nm) are pulsed for 20 ms consecutively with a delay of (20-150) ms. In each camera frame, two laser pulses produce a pair of images (green and blue) for each particle. The frame images are then split into a pair of images according to the colour. The velocity fields are obtained from these pairs of images using the cross-correlation particle image velocimetry technique. The velocity fields are measured every 0.33 s (at the camera shooting rate). For a better time resolution a video camera (2 megapixel) with a single laser is used. The damping rate (in the range of $\alpha = (0.05-0.5)$ s$^{-1}$) is estimated from the decay of the total kinetic energy $E$ after switching off the forcing at $t = t_0$: $E_t=E_{t_0}e^{-\alpha (t-t_0)}$.

\begin{figure}
\includegraphics[width=6.0 cm]{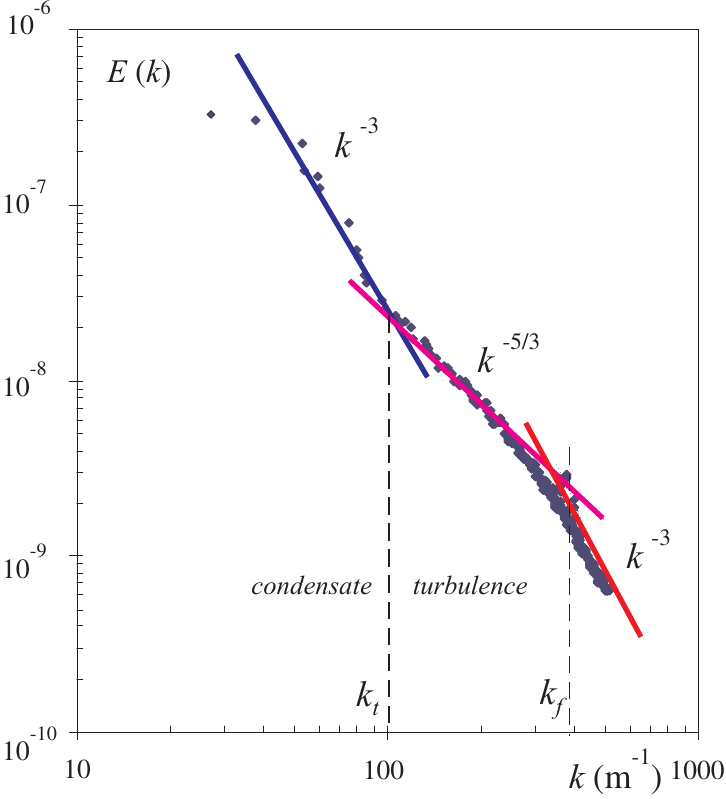}
\caption{\label{fig_1} Kinetic energy spectrum measured for the largest box $L = 0.235$ m and intermediate damping $\alpha = 0.16 s^{-1}$. The guide lines show the power laws for different spectral ranges: the $k^{-3}$ vorticity cascade, the $k^{-5/3}$ energy cascade and the $k^{-3}$ condensate range.}
\end{figure}

\begin{figure}
\includegraphics[width=6.5 cm]{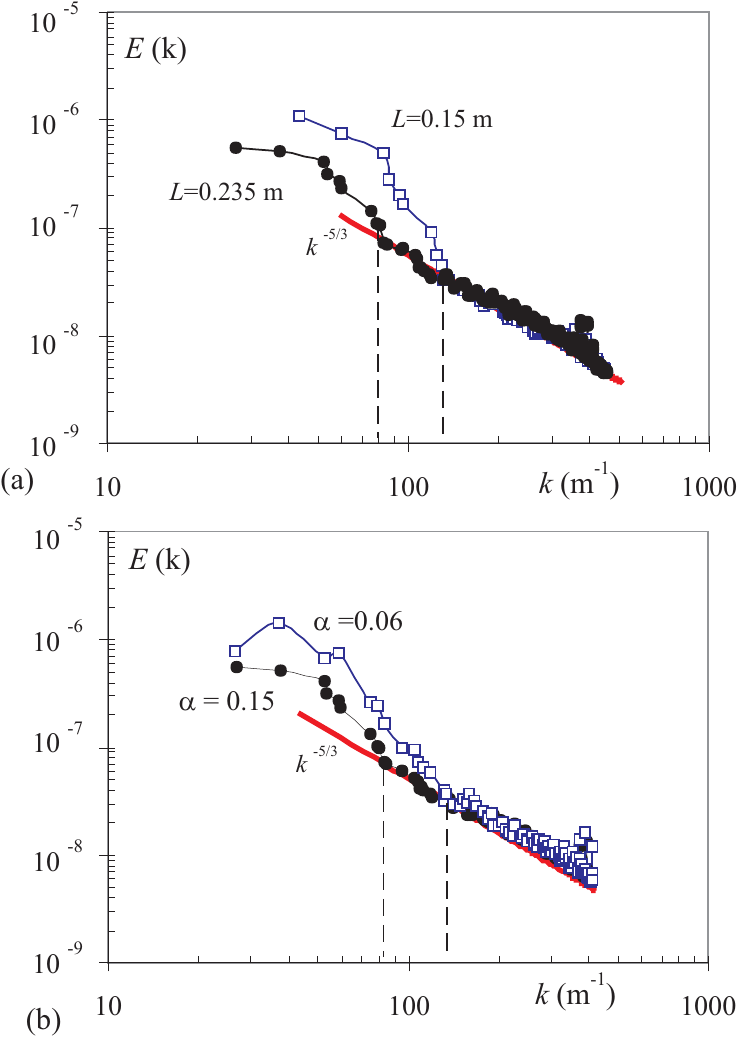}
\caption{\label{fig_2} Kinetic energy spectra (a) for different box sizes at $\alpha = 0.15$ s$^{-1}$, and (b) for different linear damping rates at $L = 0.235$ m.}
\end{figure}

Figure~\ref{fig_1} shows the energy spectrum measured in the large box $L = 0.235$ m at an intermediate damping of $\alpha = 0.16$ s$^{-1}$. A forcing scale corresponds to $k_f \approx 400$ m$^{-1}$. At $k > k_f$ , $E(k) \propto k^{-3}$, while at $k_t < k < k_f$, $E(k) \propto k^{-5/3}$. At $k < k_t \approx 80$ m$^{-1}$, in the condensate range, the spectrum is steeper and close to $k^{-3}$. Due to the presence of the condensate, the spectrum has $k^{-3}$ and $k^{-5/3}$ ranges for the large and intermediate scales respectively, similarly to the Nastrom-Gage spectrum \cite{NG84}. Spectra for different $L$ and $\alpha$ are shown in Fig.~\ref{fig_2}. At fixed $\alpha = 0.15$ s$^{-1}$, the knee of the spectrum shifts from $k_t \approx 80$ m$^{-1}$ for $L = 0.235$ m to $k_t \approx 135$ m$^{-1}$ for $L = 0.15$ m, Fig.~\ref{fig_2}(a).  When $L$ is constant, linear damping affects $k_t$ as shown in Fig.~\ref{fig_2}(b). Going from $\alpha = 0.15 s^{-1}$ to $\alpha = 0.06 s^{-1}$, changes $k_t$ from 80 $m^{-1}$ to $k_t \approx 130$ m$^{-1}$. These observations are in good qualitative agreement with Eq.~\ref{Eq:k_t}. By further reducing $L$, we can achieve a regime when $k_t \approx k_f$, and the $k^{-5/3}$ range disappears, such that the entire spectrum is $E(k) \propto k^{-3}$, both above and below $k_f$, as in \cite{Shats05}. The closer $k_t$ is to $k_f$, the more symmetric the coherent flow becomes (e.g. circular \cite{Shats05}). Therefore, we can control the shape of the spectrum and the relative strength of the condensate with respect to  turbulence.

\begin{figure}
\includegraphics[width=8.5 cm]{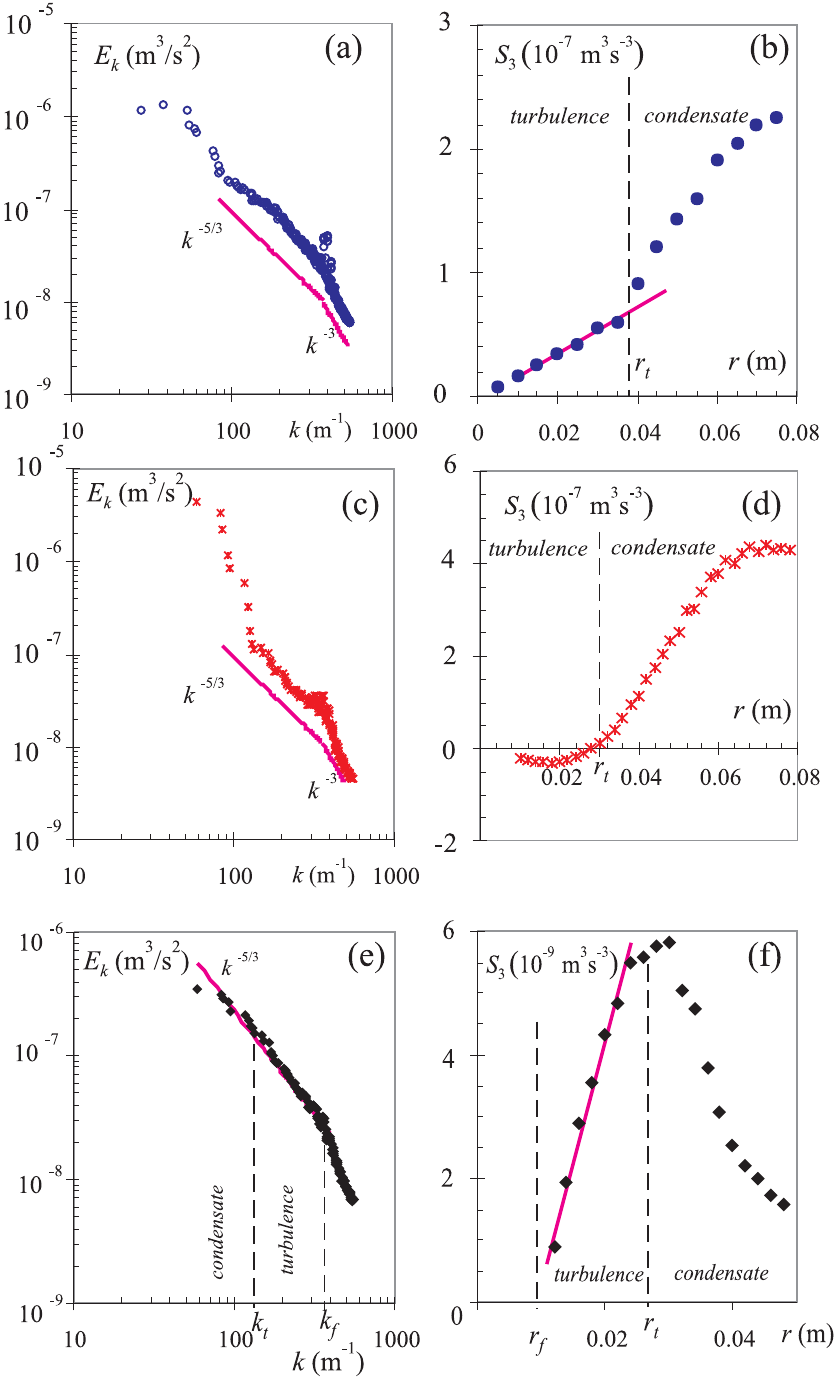}
\caption{\label{fig_3} Energy spectra $E_k(k)$ and the 3rd-order velocity moments $S_3(r)$ of turbulence with a weaker (a,b)  and stronger (c,d) condensates. After subtracting mean flow in the stronger condensate case: kinetic energy spectrum (e) and 3rd-order velocity moment (f).}
\end{figure}

We now analyse two regimes whose spectra are shown in Fig.~\ref{fig_3}(a,c). A weaker condensate of Fig.~\ref{fig_3}(a) was generated at $\alpha = 0.3$ s$^{-1}$ and $L = 0.235$ m, while a stronger condensate (Fig.~\ref{fig_3}(c)) was obtained at $\alpha = 0.15$ s$^{-1}$ and $L = 0.15$ m. The third-order velocity moments differ markedly for these two cases (Figs.~\ref{fig_3}(b,d)). For the weaker condensate case (Fig.~\ref{fig_3}(b)), $S_3$ is a linear function of the separation distance $r$ in the range of scales corresponding to the $k^{-5/3}$ spectral range. Both longitudinal $S_{3L}$ and transverse $S_{3T}$ moments are positive in the entire range of scales, in agreement with expectations for the inverse energy cascade. The slope of $S_{3}(r)$ changes at about $r = 0.04$ m. This scale corresponds to the knee in the energy spectrum at $k_t \approx 80$ m$^{-1}$.  The Kolmogorov constant can be determined as $C = E(k) \epsilon ^{-2/3} k^{5/3}$, where $\epsilon = (S_{3L}+S_{3T})/(2r)$. For the data of Fig.~\ref{fig_3}(b) at $r < \pi /k_t \approx 0.04$ m, we have $C = 5.6$, which is close to the value $C = (5.8 - 7)$ previously obtained in numerical simulations of 2D turbulence (see \cite{Boffetta00} and references therein).  At larger separations, $r > \pi/k_t$, the function $S_3(r)$ grows faster and deviates from linear.

For the stronger condensate, the spectrum scales as $E(k) \propto k^{-5/3}$ in the range $k_f > k > k_t \approx 125$ m$^{-1}$. In this case, $S_3(r)$ changes sign at $r = \pi/k_t$ (Fig.~\ref{fig_3}(d)). Such $S_3(r)$ dependence resembles the third-order structure function measured in the lower stratosphere \cite{Cho_Lindborg01}. Note that in our case all the driving comes from small scales and there is no direct cascade at all, yet $S_3$ is strongly modified compared with the weak condensate case.

The spectral condensation can be viewed as the generation of the mean flow, which can be revealed by a temporal averaging of the instantaneous velocity fields: $\bar{V}(x,y)=1/N\sum^{N}_{n=1}V(x,y,t_n)$. The power spectrum of the mean flow \cite{Chertkov07,Shats07} is close to $\bar{E}(k) \propto k^{-3}$. The velocity field contains both the mean component and turbulent velocity fluctuations: $\delta V = \delta \bar{V}+\delta \tilde{V}$. From the data of Fig.~\ref{fig_3}(c) (stronger condensate), we estimate that $\langle(\delta V)^2 \rangle$ differs from $\langle(\delta \tilde{V})^2 \rangle$ by about 20-30\%. However $\langle(\delta V)^3 \rangle$ and $\langle(\delta \tilde{V})^3 \rangle$ differ by orders of magnitude and even the sign can be different. It should be noted that signs, values and functional dependencies $S_3(r)$ vary a lot for different topologies of the condensate flows (e.g., dipole, or monopole) and also depend on the mean shear in such a flow. We observe negative and positive $S_3(r)$ in the entire range of scales, as well as the sign reversal, as in Fig.~\ref{fig_3}(d).

To recover the statistical moments of the turbulent velocity fluctuations we take $N = 350$ instantaneous velocity fields, subtract their mean flow and then compute the Fourier spectrum and the structure functions. The result for the stronger condensate of Fig.~\ref{fig_3}(c,d) is shown in Figs.~\ref{fig_3}(e,f). The subtraction of the mean restores the $k^{-5/3}$ range. In fact, the $E(k)$ scatter in Fig.~\ref{fig_3}(e) is less than in the total spectrum of Fig.~\ref{fig_3}(c). The subtraction has even more dramatic effect on $S_3$. As seen in Fig.~\ref{fig_3}(f), $S_3$ is a linear function of $r$ in the "turbulence" range. The spectral energy flux is deduced as $\epsilon = S_3/r$. The value of the Kolmogorov constant $C = E(k) \epsilon ^{-2/3} k^{5/3} \approx 7.7$ appears to be slightly higher than in the weak condensate case, but is still close to the values obtained in numerical simulations. At the border between condensate and turbulence, $r_t \approx \pi/k_t$, the $S_3(r)$ dependence changes radically, suggesting that the condensate affects the inverse energy cascade at large scales. The recovery of the linear positive $S_3(r)$ has also been observed at even stronger condensates (e.g. circular monopole flows). The Kolmogorov constant seems to become smaller at higher flow velocities and stronger shear, $C = (1 - 4)$.

It is important to note that similarity of our spectra to those of \cite{Lilly83, NG84} does not necessarily mean that $k^{-3}$ spectrum at large scales in the Earth atmosphere is also fed by the inverse cascade. To establish whether this is the case, one needs to analyze the atmospheric data in the way described here: subtract the coherent flow, recalculate the second and the third moment of fluctuations and use Eq.~\ref{Eq:k_t}. It is likely that the baroclinic (large-scale) instabilities play a role in forcing the large-scale flows. To model an extra large-scale forcing in our experiments we added a large magnet on top of the small-scale forcing (as described in \cite{Shats07}) and found that the modifications in $S_3$ are similar to those when the large-scale flow is formed via spectral condensation. The mean subtraction recovers the energy flux from small to large scales in both cases. Similarly, the mesoscale turbulence in the Earth atmosphere should be affected by the large-scale flow regardless of its origin.

Recent numerical simulations (see \cite{Brethouwer07} and
references therein) show that stratification may enforce a 3D
dynamics and the forward energy cascade. On the other hand, recent
experimental studies of  decaying turbulence suggest a strong role
of rotation in establishing a quasi-2D regime in which geostrophic
dynamics is dominant (regime of low Froude and Rossby
numbers)\cite{Praud06}. More experiments in constantly forced
turbulence are needed to better understand the competing effects
of rotation and stratification along with the complex interplay
between turbulence and waves, resonant wave-wave interactions,
etc. An ultimate answer to the question asked in the title of this
Letter can only be resolved in the atmospheric measurements.  What
we have shown here is the need to separate mean flows and
fluctuations to recover the energy flux.

\begin{acknowledgments}
The authors are grateful to V.V. Lebedev, M. Chertkov and R.E.
Ecke  for useful discussions and to D. Byrne for the help with the
data analysis. This work was partially supported by the Australian
Research Council, Israeli Science Foundation and Minerva Einstein
Center at the Weizmann Institute.
\end{acknowledgments}

\end{document}